\documentclass[graphicx,floatfix,aps,prl,twocolumn,showpacs,groupedaddress]{revtex4}
\usepackage{graphicx}
\usepackage{amsmath}

\begin{document}


\title{
  Long-range frustration in $T=0$ first-step replica-symmetry-broken
  solutions of finite-connectivity spin glasses\footnote{To be published in Journal of
Statistical Mechanics: Theory and Experiment (2007).}
}

\author{Jie Zhou, Hui Ma, and Haijun Zhou}

\affiliation{
Institute of Theoretical Physics, the Chinese Academy of Sciences, Beijing 100080, China
}
\email{zhouhj@itp.ac.cn}

\date{\today}

\begin{abstract}
  In a finite-connectivity spin-glass at the zero-temperature limit, long-range
  correlations exist among the unfrozen vertices (whose spin values being non-fixed).
  Such long-range frustrations are partially removed through the
  first-step replica-symmetry-broken (1RSB) cavity theory, but residual
  long-range frustrations may still persist in this mean-field solution. 
  By way of population dynamics, here we perform a perturbation-percolation analysis
  to calculate the magnitude of long-range frustrations in the
  1RSB solution of a given spin-glass system. We study two well-studied model systems,
  the minimal vertex-cover problem and the maximal $2$-satisfiability problem.
  This work points to a possible way of improving the zero-temperature 1RSB mean-field
  theory of spin-glasses.
\end{abstract}


\maketitle


Spin-glasses are simple models for a large class of disordered
systems with quenched randomness and frustration \cite{Fischer-Hertz-1991}.
At low temperatures, ergodicity is broken in a spin-glass
system, whose configurational phase space
splits into exponentially many ergodic domains, each of which
corresponds to a thermodynamic  state. Within each thermodynamic state, a vertex's spin
is partially frozen, but the preferred spin
orientation and the degree of preference are
vertex-dependent.
Mean-field theories for spin-glasses on random graphs
of finite connectivity \cite{Monasson-1998,Mezard-Parisi-2001} have attracted
a lot of research interest in recent years.
The work of M{\'{e}}zard and
Parisi \cite{Mezard-Parisi-2001} combines the classical Bethe-Peierls
approximation (BA)
\cite{Bethe-1935} with the exponential proliferation
of thermodynamic states.
It is a first-step replica-symmetry-broken
(1RSB) cavity solution for
finite-connectivity spin-glasses \cite{Mezard-Parisi-2001}.
The zero-temperature limit of this mean-field theory
\cite{Mezard-Parisi-2003} has found important applications in computer
science and information theory \cite{Mezard-etal-2002,Mezard-Zecchina-2002}.

The essence of the BA is  assuming statistical independence among the vertices in
the nearest-neighbor set $\partial i$ of any given vertex $i$ in the cavity graph
where $i$ is removed. Very recently, the cavity approximation and loop expansion method
\cite{Montanari-Rizzo-2005,Parisi-Slanina-2006,Chertkov-Chernyak-2006a,Chertkov-Chernyak-2006b}
were developed, which calculate the  statistical correlations
among cavity vertices in $\partial i$. These approaches
at the present stage work only in the ergodic high-temperature paramagnetic phase.
Another theoretical approach of going beyond the BA, which
works in the spin-glass phase at the other limit of temperature
$T \to 0$ \cite{Zhou-2005a,Zhou-2005b}, is to consider long-range
frustrations among the unfrozen cavity vertices
in the set $\partial i$. At the $T \to 0$ limit,
a vertex $i$ of a spin-glass system either takes the same spin value  in all the relevant
configurations which contribute to the free energy of the
system, or is unfrozen and takes different spin values
in different configurations.
With respect to a pre-specified value $\sigma^* \in \pm 1$, an unfrozen
vertex $i$ may be type-I or type-II unfrozen, depending on whether
or not the fixation of $\sigma_i$ to $\sigma_i= \sigma^*$  leads to the
fixation of the spin values of a {\em finite fraction} of
all the other unfrozen vertices \cite{Zhou-2005a,Zhou-2005b}.
All the type-I unfrozen vertices in a spin-glass system are strongly correlated,
no matter how far apart they are separated from each other.
Such strong correlations are referred to as long-range frustrations.

The present work represents a first step in the on-going effort of
integrating the physical idea of long-range frustration into the 1RSB
mean-field cavity approach of spin-glasses.
Long-range frustration is associated with ergodicity-breaking:
type-I unfrozen vertices are absent in the  paramagnetic
phase. Even in the spin-glass phase, long-range frustration should vanish
within each ergodic subdomain of the configurational
space.  However, the configurational space of a spin-glass system
may be organized far more complex than what is assumed in the 1RSB mean-field
theory \cite{Mezard-Parisi-2003}.
For example, a macroscopic state of the 1RSB solution
may actually be a merge of several true thermodynamic states
(referred to as
 the second kind or type II instabilities in Ref.~\cite{Rivoire-etal-2004} and
Ref.~\cite{Montanari-etal-2004}).
Therefore, it may be possible that
type-I unfrozen vertices still exist
within a macroscopic state of the 1RSB mean-field solution. Here this
possibility is checked by a quantitative percolation analysis (by population
dynamics) of spin-flip perturbations.
We work on two concrete models,
the minimal vertex-cover problem \cite{Hartmann-Weigt-2003} and
the maximal $2$-satisfiability problem \cite{Fernandez-2001}; we find
that  residual long-range frustrations exist in the former system but
are absent in the later system. This work suggests a possible way of improving
the zero-temperature 1RSB mean-field theory of spin-glasses.
We  also discuss possible other extensions of the present theoretical framework.
The existence of residual long-range frustrations is
a signature of the instability of the $T=0$ 1RSB mean-field solution toward
further steps of replica-symmetry-breaking
\cite{Rivoire-etal-2004,Montanari-etal-2004,Montanari-RicciTersenghi-2003,Gardner-1985}.
It is not clear, however, whether the onset of long-range frustration is a necessary
condition for the 1RSB macroscopic state to be non-ergodic (i.e., made of smaller
sub-states).


{\em Percolation analysis}.---Type-I unfrozen vertices
form a giant cluster in a graph, their existence therefore can be
detected by a percolation analysis. (The propagation of
correlations among type-I unfrozen variables is close in spirit to the notion of
bug proliferation of Ref.~\cite{Mertens-etal-2006}.)
 To introduce the basic
methodology of percolation analysis by population dynamics, let us as an example
calculate the fraction $q$ of vertices that
are in the giant component of a random Poissonian graph of mean vertex-degree
$c$ \cite{Bollobas-1985}. We note that
a vertex $i$ is in the giant component if at least one vertex in its nearest-neighbor
set $\partial i$ is in the giant component of the cavity graph where $i$ is removed.
We  construct a population of binary elements, each of which is either
$1$ (in the giant component) or $0$ (not in the giant component). At each elementary
update, $k$ elements are randomly chosen from the population ($k$ being a
random integer governed by the Poisson distribution of mean $c$); then
a randomly chosen element of the population is set to be $0$ if all these $k$
(input) elements are  zero, otherwise it is set to be $1$.
The giant component size $q$ is estimated to be
the fraction of $1$'s in the whole population. We found that this value
is identical to the value predicted by the well-known formula $q=1-e^{-c q}$
\cite{Bollobas-1985}.  In what follows  we analyze
long-range frustrations in spin-glasses using this methodology. We demonstrate our
approach by first working on the minimal vertex-cover problem.


{\em The minimal vertex-cover}.---For a given graph ${\cal G}$  of $N$ vertices
and $M$ edges $(i,j)$ between pairs of vertices $i$ and $j$,
a minimal vertex-cover is a spin pattern $\{ \sigma_i \in \pm 1 \}$  which satisfies the
constraint
\begin{equation}
  \label{eq:mvc-constraint}
  \prod\limits_{(i,j)\in {\cal G} }\Bigl[ 1- \frac{(1+\sigma_i)(1+\sigma_j)}{4} \Bigr] \equiv 1 \ ,
\end{equation}
and which  minimizes the total energy
\begin{equation}
  \label{eq:mvc-energy}
  E \bigl(\sigma_1,  \ldots,  \sigma_N \bigr) = \sum\limits_{i=1}^{N} \frac{  1 - \sigma_i}{2 } \ .
\end{equation}
A minimal vertex-cover for a given graph ${\cal G}$ is a spin pattern with the maximal number of
uncovered ($+1$) vertices, while for each edge of the graph at least one of its two
end vertices is covered ($\sigma=-1$).
The mean-field 1RSB cavity solution of the minimal vertex-cover 
problem on finite-connectivity random graphs was
reported in Ref.~\cite{Weigt-Zhou-2006}. When the mean vertex degree $c$ of the random graph satisfies
$c > 2.7183$, the system is in the spin-glass phase.
In this spin-glass phase, the 1RSB theory assumes that in the limit of
graph size $N \to \infty $, there are
exponentially many local optimal vertex-covers
for the constrained system.
Each local optimal pattern is stable with respect to any perturbation
which flips a finite number of spins.
A macroscopic state $\alpha$ of the 1RSB mean-field solution contains a set of
local optimal patterns of the same energy $E_\alpha$. Two patterns in the same macroscopic
state are assumed to be similar to each other.
At $T=0$, the total grand free energy $G$ of the system
can be defined by the equation
\begin{equation}
  \label{eq:grand-free-energy}
  \exp\bigl( - y G \bigr) = \sum\limits_{\alpha} \exp\bigl( - y E_\alpha \bigr) \ ,
\end{equation}
where the summation is over all the macroscopic states of the system;
$y$ is a re-weighting parameter \cite{Mezard-Parisi-2003}, whose value is
chosen such that $G(y)$ attains maximality \cite{Zhou-2007b}.

A vertex $i$ is referred to as being
positively (negatively) frozen in macroscopic state $\alpha$ if
its spin value  is positive (negative) in all the configurations
of this macroscopic state. This spin value freezing is caused by
energy minimization under the constraint Eq.~(\ref{eq:mvc-constraint}).
Vertex $i$ is said to be unfrozen in macroscopic
state $\alpha$ if it is neither positively nor negatively frozen.
If in the cavity graph ${\cal G}\backslash i$  all the vertices in
the nearest-neighbor set $\partial i$ of vertex $i$ are negatively frozen 
in state $\alpha$, then $i$ will be positively frozen in macroscopic state 
$\alpha$ of the full graph
${\cal G}$; on the other hand,  if two or more vertices 
in $\partial i$ are positively frozen in state $\alpha$ of ${\cal G}\backslash i$,
then $i$ will be negatively frozen in state $\alpha$ of ${\cal G}$.
However, the remaining situations are tricky.
As an example, consider the case in which
all vertices in the set $\partial i$ are unfrozen in state $\alpha$ of the
cavity graph ${\cal G}\backslash i$. When vertex $i$ is added to the system,
all the vertices in $\partial i$ {\em should} take the minus spin value so that
vertex $i$ will no longer need to be covered.
But this is possible only when all the unfrozen cavity
vertices in $\partial i$ are able to take the minus spin value {\em simultaneously}. If
all these cavity vertices are type-II unfrozen, there is no problem.
But if two (say $j$ and $k$)
or more of these vertices are type-I unfrozen (with respect to the
minus spin value, of cause), it may be that flipping $\sigma_j$ to $\sigma_j=-1$ will
force the spin of vertex $k$ to be $\sigma_k=+1$! This is because when a type-I
unfrozen vertex $j$  is fixed to
$\sigma_j=-1$, eventually a percolating cluster of other unfrozen vertices
in the whole system will also have their spins be fixed (either positively
or negatively); all the other type-I unfrozen vertices are in this
giant cluster \cite{Zhou-2005a}.

In the mean-field 1RSB
cavity solution \cite{Weigt-Zhou-2006} all the unfrozen vertices are
assumed to be type-II, and no long-range frustration effect is
considered.  Here let us
assume that initially  some of the vertices in each macroscopic state are
type-I unfrozen.  We will then check whether
the fraction of type-I unfrozen vertices will shrink to
zero in the 1RSB population dynamics. We will focus on whether type-I
unfrozen vertices persist but not on their energetic effects.
In other words, we assume that the type-I unfrozen
cavity vertices in the nearest-neighbor set $\partial i$ of each vertex $i$
can take the minus spin simultaneously. Under this simplification,
let us consider a cavity vertex $j$ which is connected by an edge $(i,j)$ to
a vertex $i$ of the graph $G$. In the cavity graph $G\backslash i$ in which
vertex $i$ is being removed, the cavity vertex $j$ will be type-I unfrozen if: (1)
only one of the vertices (say $k$) 
in the set $\partial j \backslash i$ is positively frozen
in the graph ${\cal G}\backslash i,j$, and (2) this vertex $k$ is itself
connected to one or more type-I unfrozen vertices of the cavity
graph ${\cal G}\backslash i,j,k$ \cite{Zhou-2005a}.
Let us denote by $\hat{\pi}_{j\to i}^{0}$ the fraction of macroscopic states
in which vertex $j$ is positively frozen and
none of its nearest-neighbors is type-I unfrozen in the
cavity graph ${\cal G}\backslash i,j$.
Similarly, $\tilde{\pi}_{j\to i}^{0}$ is the fraction of states in which
$j$ is positively frozen and some of its nearest neighbors is type-I unfrozen
in ${\cal G}\backslash i,j$; $\hat{\pi}_{j\to i}^{*}$
is the fraction of  states in which
$j$ is type-II unfrozen in graph ${\cal G}\backslash i$;
and $\tilde{\pi}_{j\to i}^{*}$ is the fraction of  states
in which $j$ is type-I unfrozen in graph ${\cal G}\backslash i$. 
Following the 1RSB cavity approach
\cite{Weigt-Zhou-2006} we can write down the following iterative equations for a
Poissonian random graph:

\begin{widetext}
\begin{eqnarray}
  \hat{\pi}_{j\to i}^{0} &=& \frac{ \prod\limits_{k\in \partial j\backslash i} 
    [1- \hat{\pi}_{k\to j}^{0}-\tilde{\pi}_{k\to j}^{0}-
    \tilde{\pi}_{k\to j}^{*} ] }{
    e^{-y} + (1- e^{-y} ) \prod\limits_{k \in \partial j\backslash i}
    [1- \hat{\pi}_{k\to j}^{(0)}-\tilde{\pi}_{k\to j}^{(0)}] }
  \ ,
\nonumber \\
\tilde{\pi}_{j\to i}^{0} &=&
\frac{  \prod\limits_{k \in \partial j \backslash i} [1- \hat{\pi}_{k\to j}^{0}
  -\tilde{\pi}_{k\to j}^{0}] -
  \prod\limits_{k\in \partial j \backslash i} [1- \hat{\pi}_{k\to j}^{0}
  -\tilde{\pi}_{k\to j}^{0}-\tilde{\pi}_{k\to j}^{*} ] }{
  e^{-y} + (1- e^{-y} ) \prod\limits_{k\in \partial j\backslash i}
  [1- \hat{\pi}_{k\to j}^{(0)}-\tilde{\pi}_{k\to j}^{(0)}] } \ ,
\nonumber \\
\hat{\pi}_{j\to i}^{*} &=&
\frac{ e^{-y} \sum\limits_{ k\in \partial j \backslash i} 
  \hat{\pi}_{k\to j}^{0} \prod\limits_{l\in \partial j \backslash i,k }
  [1- \hat{\pi}_{l\to j}^{0}-\tilde{\pi}_{l\to j}^{0}] }{
  e^{-y} + (1- e^{-y} ) \prod\limits_{k\in \partial j\backslash i} 
  [1- \hat{\pi}_{k\to j}^{(0)}-\tilde{\pi}_{k\to j}^{(0)}] } \ ,
\nonumber \\
\tilde{\pi}_{j\to i}^{*} &=& \frac{ e^{-y} \sum\limits_{k\in \partial j\backslash i}
  \tilde{\pi}_{k\to j}^{0} \prod\limits_{l\in \partial j \backslash  i,k }
  [1- \hat{\pi}_{l\to j}^{0}-\tilde{\pi}_{l\to j}^{0}] }{
  e^{-y} + (1- e^{-y} ) \prod\limits_{k\in \partial j\backslash i}
  [1- \hat{\pi}_{k\to j}^{(0)}-\tilde{\pi}_{k\to j}^{(0)}] }
\ .
\label{eq:pi}
\end{eqnarray}

\end{widetext}

Two steady-state distributions $P(\hat{\pi}_{j\to i}^{0},
\tilde{\pi}_{j\to i}^{0}, \hat{\pi}_{j\to i}^{*},
\tilde{\pi}_{j\to i}^{*})$ and  $P(\hat{\pi}_{i\to j}^{0},
\tilde{\pi}_{i\to j}^{0}, \hat{\pi}_{i\to j}^{*},
\tilde{\pi}_{i\to j}^{*})$
on all edges $(i,j)$ of the graph can be obtained by population dynamics simulations
\cite{Weigt-Zhou-2006}.
An array of ${\cal N}=10^6$ vectors ${\bf \pi}_{j\to i}
= \{ \hat{\pi}_{j\to i}^{0}, \tilde{\pi}_{j\to i}^{0},
\hat{\pi}_{j\to i}^{*}, \tilde{\pi}_{j\to i}^{*} \} $ is constructed;
elements of this array are then updated according to Eq.~(\ref{eq:pi}).
At given value of the re-weighting parameter $y$
the grand free-energy density of the system is calculated \cite{Weigt-Zhou-2006},
as well as the following long-range order parameter $R_{j\to i}$ for each directed
edge $j\to i$:
\begin{equation}
  \label{eq:long-range-order-parameter}
  R_{j\to i} \equiv \frac{ \tilde{\pi}_{j\to i}^{*} }{ 
    ( \hat{\pi}_{j\to i}^{*} + \tilde{\pi}_{j\to i}^{*} ) } \ .
\end{equation}
$R_{j\to i}$ measures the probability of an unfrozen cavity vertex $j$ being actually
type-I unfrozen.

\begin{figure}[htb]
 \includegraphics[width=0.9\linewidth]{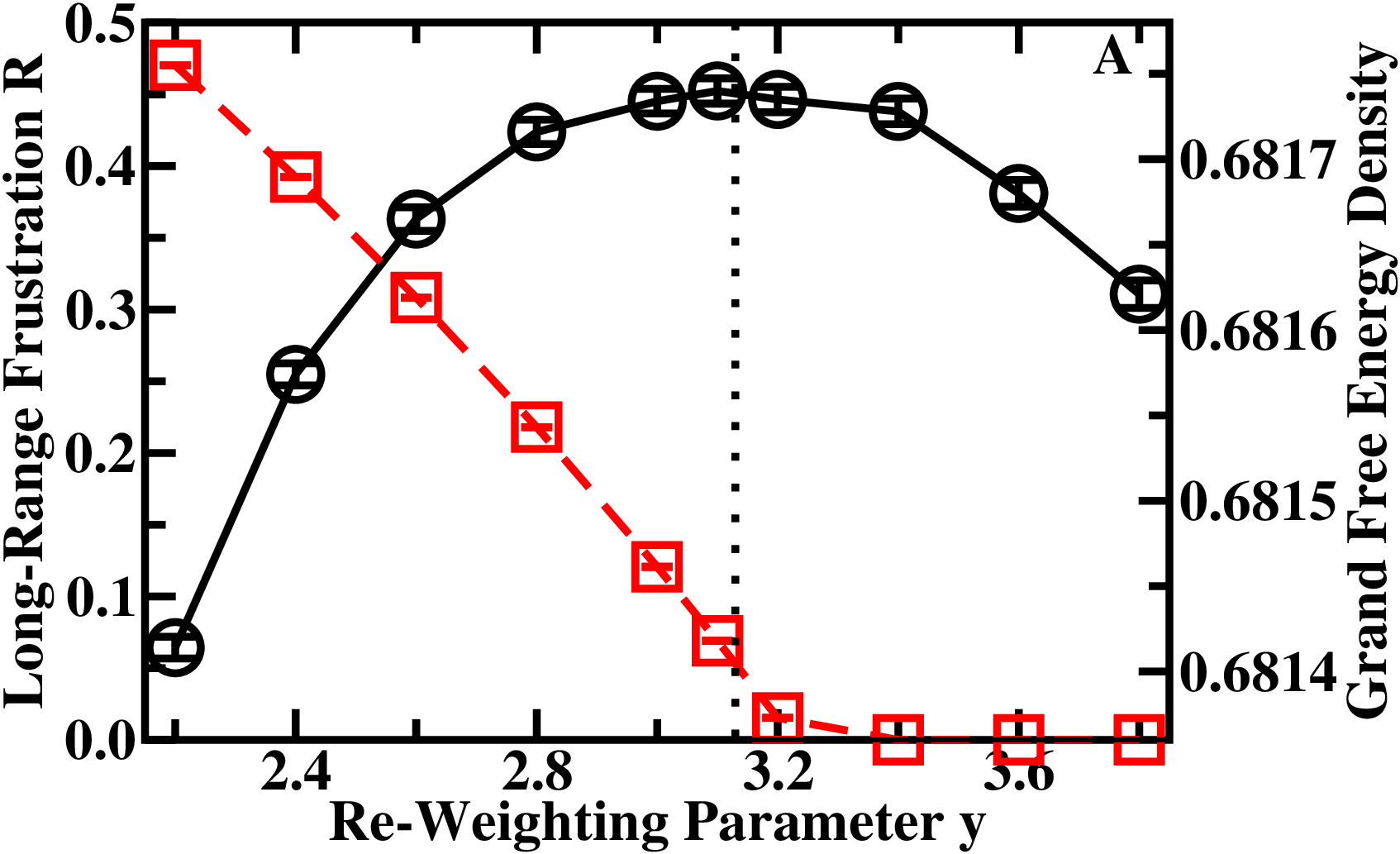}
    
  \vskip 1.0cm
  \includegraphics[width=0.9\linewidth]{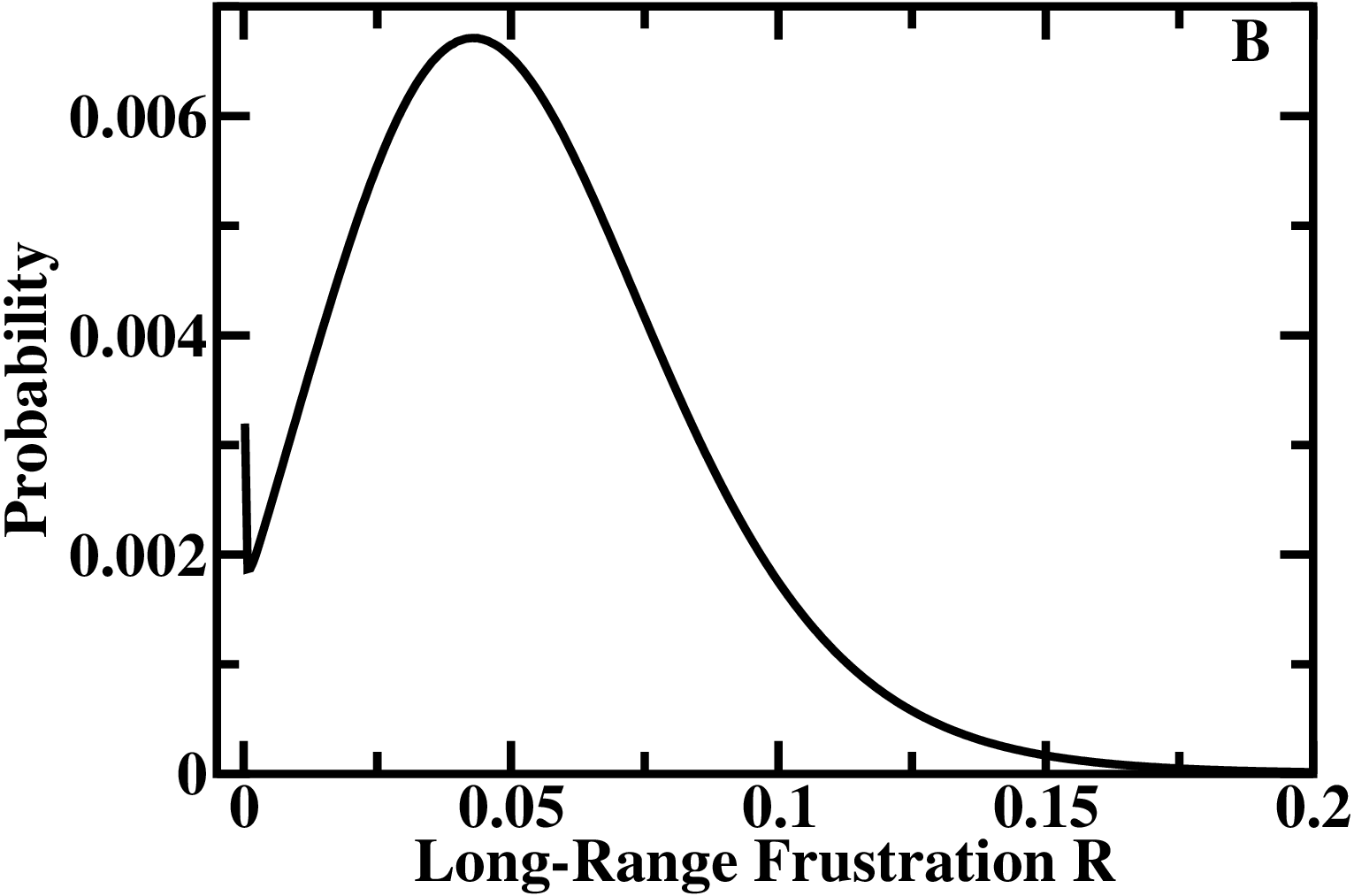}
  \vskip 1.0cm
  \caption{
    \label{fig:mvc}
    Long-range frustration in the minimal vertex-cover problem on an ensemble
    of random Poissonian graph of mean vertex degree $c=10$.
    (A) Grand free energy density (circles) and mean long-range frustration 
    $R$ (squares). The vertical dotted line indicates the point of
    $y=3.1297$  where the grand free energy density is maximal.
    (B) The distribution of the $R_{j\to i}$ values at $y=3.1297$. 
  }
\end{figure}

For an ensemble of random Poissonian graph with mean vertex degree $c=10$, the
population dynamics results are shown in Fig.~\ref{fig:mvc}. At $y=y^*=3.130(8)$ 
the grand free-energy density reaches maximum, which
corresponds to the ground-state energy density of the system. The mean long-range
frustration order-parameter $R=(1/{\cal N}) \sum R_{j\to i}$ decreases with the
re-weighting parameter $y$ and becomes zero for $y > 3.25$. Most importantly,
at $y=y^*$ the value of $R$ is positive ($R=0.0534$).
Figure~\ref{fig:mvc}B shows the distribution
of the $R_{j\to i}$ values at $y=y^*$. This distribution has a peak at $R_{j\to i}=0$ and
another peak at
$R_{j\to i} \approx 0.043$. On average an unfrozen vertex is type-I unfrozen in about
five percent of the macroscopic states.
We therefore conclude that in the 1RSB mean-field treatment of the minimal
vertex-cover problem there exist still residual long-range frustrations.
We have also checked that the above qualitative conclusion
holds also for other values the mean vertex degrees $c > 2.7183$.
For example, at $c=5$, the mean long-range frustration order
parameter is $R=0.0405$, slightly below the value for $c=10$.

Does long-range frustration exist in the 1RSB mean-field cavity solutions
of  all finite-connectivity
spin-glass models with two-body interactions? The following example
suggests that this is not necessarily true.


{\em The maximal $2$-satisfiability problem}.---On a
random factor graph \cite{Kschischang-etal-2001} ${\cal G}$ with $N$ variable nodes
$i$ and $M= \alpha N$ function nodes $a$, each of which connects to two randomly chosen 
variable nodes $i$ and $j$, the $2$-SAT energy function is defined as
\begin{equation}
\label{eq:2sat-energy}
E( \sigma_1, \ldots, \sigma_N ) = \sum\limits_{a \in {\cal G}} \frac{ (1- J_a^i \sigma_i)
(1-J_a^j \sigma_j) }{ 4} \ .
\end{equation}
In Eq.~(\ref{eq:2sat-energy}) the quenched coupling constant  $J_a^i$ of the edge
between the function node $a$ and variable node $i$ is equal to $+1$ or
$-1$ with equal probability. The Max-2SAT problem consists of finding
a binary spin pattern which minimizes the configurational energy
Eq.~(\ref{eq:2sat-energy}). For a random factor graph, it is well known that,
the ground-state energy of the system is zero when $\alpha < 1$
\cite{Fernandez-2001}; it becomes positive for $\alpha >1$, and long-range
frustration builds up in the Max-2SAT
when $\alpha > 4.4588$ \cite{Zhou-2005b}.

\begin{figure}[htb]
    \includegraphics[width=0.9\linewidth]{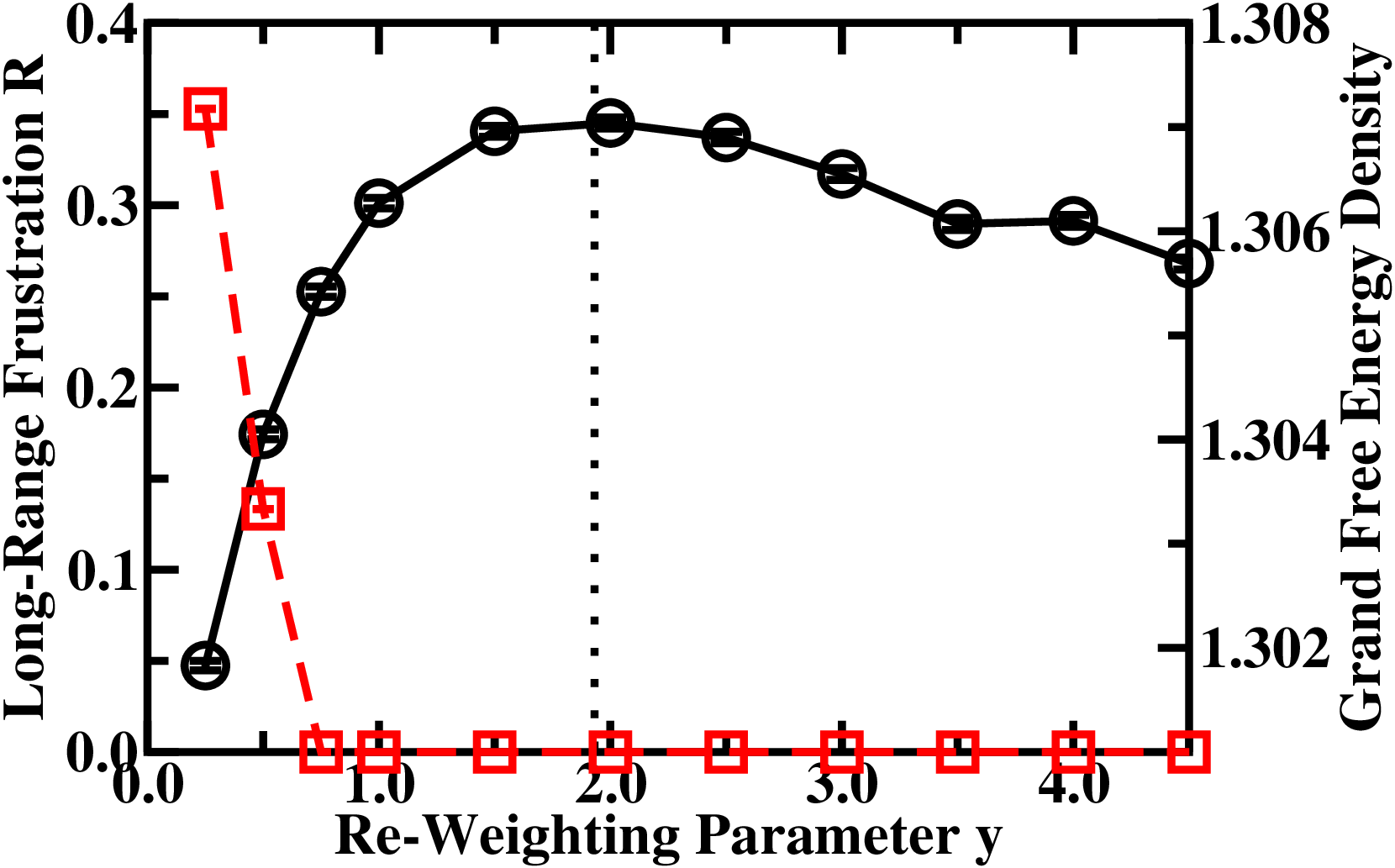}
  \caption{
    \label{fig:2sat}
    Grand free energy density (circles) and mean long-range frustration
    $R$ (squares) for the random 2-SAT problem with $\alpha=10$.
    The vertical dotted line indicates the point of $y=1.932$.
  }
\end{figure}

We have applied the perturbation-percolation analysis on the 1RSB solution of
the Max-2SAT problem defined on an ensemble of random factor graphs 
of $\alpha=10$. As shown in Fig.~\ref{fig:2sat},
the grand free-energy density of the system reaches maximality at
$y=1.932$, but the mean long-range frustration order parameter
$R$ already drops to zero at $y \approx 0.75$. (For $\alpha=20$ similar
results are obtained.) 
Consequently, the present percolation analysis suggests that the 1RSB
mean-field solution of the Max-2SAT problem is free of any residual
long-range frustrations.
Whether or not long-range frustrations persist
in the mean-field 1RSB solution therefore also depends on the particular
system under study.

By passing we note that the 1RSB mean-field theory reports a
ground-state energy density of $1.307040(2)$. Interestingly,
this value is just located
at the middle between the predicted lower- and upper-bound of
$1.301378$ and $1.311933$ \cite{Zhou-2005b}.


{\em Conclusion and discussion}.---In summary, in this paper we have
investigated by population dynamics the possibility of long-range
correlations among the unfrozen vertices in the zero-temperature 
first-step replica-symmetry-breaking solutions of finite-connectivity
spin-glasses. The perturbation-percolation analysis of
this work demonstrated that, residual long-range frustrations may still
persist in the mean-field 1RSB cavity solutions. The present theoretical
approach is also able to give a quantitative estimate of the magnitude
of long-range frustrations in the 1RSB mean-field theory.

At the present stage we are interested in the existence or not of type-I
unfrozen vertices in the 1RSB mean-field theory, we have not yet
actually calculated the energetic effects of type-I unfrozen vertices.
We hope to return to this point in later publications. If the mean
long-range frustration order parameter is positive, energetic effects of
long-range frustrations can be included into the present population
dynamics approach, and a better estimate of the ground-state energy
density can be obtained.
In the minimal vertex-cover problem, given the fact that
the mean long-range frustration order parameter $R$ is of the order
of $10^{-2}$ in the 1RSB solution, the
correction to the ground-state energy of the system might turn out to
be very small.

The presence of long-range correlations may reflect the instability of the 1RSB mean-field
solution. Each macroscopic state of the 1RSB solution may be non-ergodic and can be
further divided into a group of sub-states.
A procedure of stability analysis for zero-temperature 1RSB mean-field solutions has
been outlined in Refs.~\cite{Rivoire-etal-2004,Montanari-etal-2004}.
It is of interest to apply this analysis
to spin-glass problems with two-body interactions and check whether or not
its qualitative conclusions
are consistent with the results obtained by the long-range frustration analysis.
The estimated mean long-range frustration value $R$ should depends on
the type of ansatz (1RSB, 2RSB,\ldots) used for the computation. When a macroscopic state
of the 1RSB solution further breaks into sub-macroscopic states (2RSB),
a type-I unfrozen vertex
in the macroscopic state of the 1RSB solution may become frozen in the
daughter sub-macroscopic states of the 2RSB solution. The mean long-range frustration value $R$
therefore will decreases as higher steps of RSB are assumed.
The present work also suggested that, for some spin-glass systems,
long-range frustrations may be absent in the
1RSB mean-field cavity solutions.
For these systems, maybe the mean-field
1RSB cavity theory is already enough to describe their equilibrium properties
at temperature $T\to 0$. Higher levels of replica-symmetry-breaking
may not be needed any more.
However, it is not known whether the existence of long-range frustration is a necessary
condition for the zero-temperature 1RSB mean-field solution to be unstable.
 This point should be checked seriously.

Both the earlier \cite{Zhou-2005a,Zhou-2005b} and the present work 
study the statistical mechanical properties of finite-connectivity spin-glasses
at the limit of zero temperature.  When the temperature $T$ is low but 
still positive, long-range correlations may also exist in mean-field solutions of 
spin-glass systems. However, at finite temperatures we can not make the
distinction between frozen and unfrozen vertices. This is because all the vertices are
unfrozen. How to extend the present theoretical approach to the case of
positive temperatures is an interesting and challenging problem.


JZ and HM are grateful to Prof.~Zhong-Can Ou-Yang for support. Part of the numerical 
simulations were performed at the PC clusters of the State Key Laboratory for Scientific and
Engineering Computing, CAS, Beijing.



\begin{thebibliography}{10}

\bibitem{Fischer-Hertz-1991}
Fischer, K.H., Hertz, J.A.:
\newblock Spin Glasses.
\newblock Cambridge Univ. Press, Cambridge (1991)

\bibitem{Monasson-1998}
Monasson, R.:
\newblock Optimization problems and replica symmetry breaking in finite
  connectivity spin glasses.
\newblock J. Phys. A: Math. Gen. \textbf{31} (1998)  513--529

\bibitem{Mezard-Parisi-2001}
M{\'{e}}zard, M., Parisi, G.:
\newblock The bethe lattice spin glass revisited.
\newblock Eur. Phys. J. B \textbf{20} (2001)  217--233

\bibitem{Bethe-1935}
Bethe, H.A.:
\newblock Statistical theory of superlattices.
\newblock Proc. R. Soc. London A \textbf{150} (1935)  552--575

\bibitem{Mezard-Parisi-2003}
M{\'{e}}zard, M., Parisi, G.:
\newblock The cavity method at zero temperature.
\newblock J. Stat. Phys. \textbf{111} (2003)  1--34

\bibitem{Mezard-etal-2002}
M{\'{e}}zard, M., Parisi, G., Zecchina, R.:
\newblock Analytic and algorithmic solution of random satisfiability problems.
\newblock Science \textbf{297} (2002)  812--815

\bibitem{Mezard-Zecchina-2002}
M{\'{e}}zard, M., Zecchina, R.:
\newblock The random k-satisfiability problem: from an analytic solution to an
  efficient algorithm.
\newblock Phys. Rev. E \textbf{66} (2002)  056126

\bibitem{Montanari-Rizzo-2005}
Montanari, A., Rizzo, T.:
\newblock How to compute loop corrections to bethe approximation.
\newblock J. Stat. Mech.: Theo. Exp. (2005)  P10011

\bibitem{Parisi-Slanina-2006}
Parisi, G., Slanina, F.:
\newblock Loop expansion around the bethe-peierls approximation for lattice
  models.
\newblock J. Stat. Mech.: Theo. Exp. (2006)  L02003

\bibitem{Chertkov-Chernyak-2006a}
Chertkov, M., Chernyak, V.Y.:
\newblock Loop calculus in statistical physics and information science.
\newblock Phys. Rev. E \textbf{73} (2006)  065102(R)

\bibitem{Chertkov-Chernyak-2006b}
Chertkov, M., Chernyak, V.Y.:
\newblock Loop series for discrete statistical models on graphs.
\newblock J. Stat. Mech.: Theor. Exp. (2006)  P06009

\bibitem{Zhou-2005a}
Zhou, H.:
\newblock Long-range frustration in a spin-glass model of the vertex-cover
  problem.
\newblock Phys. Rev. Lett. \textbf{94} (2005)  217203

\bibitem{Zhou-2005b}
Zhou, H.:
\newblock Long-range frustration in finite connectivity spin glasses: a
  mean-field theory and its application to the random $k$-satisfiability
  problem.
\newblock New J. Phys. \textbf{7} (2005)  123

\bibitem{Rivoire-etal-2004}
Rivoire, O., Biroli, G., Martin, O.C., M{\'{e}}zard, M.:
\newblock Glass models on bethe lattice.
\newblock Eur. Phys. J. B \textbf{37} (2004)  55--78

\bibitem{Montanari-etal-2004}
Montanari, A., Parisi, G., {Ricci-Tersenghi}, F.:
\newblock Instability of one-step replica-symmetry-broken phase in
  satisfiability problems.
\newblock J. Phys. A: Math. Gen. \textbf{37} (2004)  2073--2091

\bibitem{Hartmann-Weigt-2003}
Hartmann, A.K., Weigt, M.:
\newblock Statistical mechanics of the vertex-cover problem.
\newblock J. Phys. A: Math. Gen. \textbf{36} (2003)  11069--11093

\bibitem{Fernandez-2001}
{Fernandez de la Vega}, W.:
\newblock Random 2-sat: results and problems.
\newblock Theor. Comput. Sci. \textbf{265} (2001)  131--146

\bibitem{Montanari-RicciTersenghi-2003}
Montanari, A., {Ricci-Tersenghi}, F.:
\newblock On the nature of the low-temperature phase in discontinuous
  mean-field spin glasses.
\newblock Eur. Phys. J. B \textbf{33} (2003)  339--346

\bibitem{Gardner-1985}
Gardner, E.:
\newblock Spin glasses with $p$-spin interactions.
\newblock Nucl. Phys. B \textbf{257 [FS14]} (1985)  747--765

\bibitem{Mertens-etal-2006}
Mertens, S., M{\'{e}}zard, M., Zecchina, R.:
\newblock Threshold values of random $k$-sat from the cavity method.
\newblock Rand. Struct. Algorithms \textbf{28} (2006)  340--373

\bibitem{Bollobas-1985}
Bollob{\'{a}}s, B.:
\newblock Random Graphs.
\newblock Academic Press, Landon (1985)

\bibitem{Weigt-Zhou-2006}
Weigt, M., Zhou, H.:
\newblock Message passing for vertex covers.
\newblock Phys. Rev. E \textbf{74} (2006)  046110

\bibitem{Zhou-2007b}
Zhou, H.:
\newblock Boltzmann distribution of free energies in a finite-connectivity
  spin-glass system and the cavity approach.
\newblock Frontiers of Physics in China \textbf{2} (2007)  238--250

\bibitem{Kschischang-etal-2001}
Kschischang, F.R., Frey, B.J., Loeliger, H.A.:
\newblock Factor graphs and the sum-product algorithm.
\newblock IEEE Trans. Infor. Theor. \textbf{47} (2001)  498--519

\end{thebibliography}

\end{document}